\documentclass[12pt]{article}
\newcommand {\kpipi} {$K^0_S \rightarrow \pi^+\pi^-$}

\newcommand {\numucc} {$\nu_\mu$CC}
\newcommand {\anumucc} {$\bar{\nu}_\mu$CC}
\newcommand {\kzepro} {$K^0_S p$}
\newcommand {\mkzepro} {$m(K^0_S p)$}

\begin {document}
\title
{Evidence for formation of a narrow $K^0_S p$ resonance with mass
near 1533 MeV in neutrino interactions}
\author { 
A.E. Asratyan\thanks{Corresponding author. Tel.: +7-095-237-0079;
fax: +7-095-127-08-37; E-mail address: \mbox{asratyan@vitep1.itep.ru.}},
A.G. Dolgolenko, and
M.A. Kubantsev\thanks{Now at Department of Physics and Astronomy,
Northwestern University, Evanston, IL60208, USA.}\\
\normalsize {\it Institute of Theoretical and Experimental Physics,}\\
\normalsize {\it B. Cheremushkinskaya 25, Moscow 117259, Russia}\\
}                                          % authors
%  \date {\today}
\date{Submitted to Yad. Fiz. (Phys. At. Nucl.)}
\maketitle

\begin{abstract}
A narrow baryon resonance is observed in invariant mass of the \kzepro\ 
system formed in neutrino and antineutrino collisions with nuclei. The
mass of the resonance is estimated as $1533 \pm 5$ MeV. The observed
width is less than 20 MeV, and is compatible with being entirely
due to experimental resolution. The statistical
significance of the signal is near 6.7 standard deviations. The analysis 
is based on the data obtained in past neutrino experiments with big 
bubble chambers: WA21, WA25, WA59, E180, and E632.
\end{abstract}

\newpage
     A narrow baryon resonance with mass near 1540 MeV and unnatural
(positive) strangeness has been recently detected in the $K^+ n$ 
system formed in the reaction $\gamma n \rightarrow K^+ K^- n$ on
Carbon \cite{nakano} and in the $K^0 p$ system from the charge-exchange 
reaction $K^+ n \rightarrow K^0 p$ in low-energy $K^+$Xe 
collisions \cite{xenon}. Similar observations have since been reported
by two other photoproduction experiments \cite{clas, saphir}.
This object, referred as the $\Theta^+(1540)$, is tentatively
interpreted as the lightest member of an antidecuplet of pentaquark
baryons, as predicted some time ago in the framework of the chiral
soliton model \cite{theory}. This paper reports on a search for formation
of the $\Theta^+$ baryon in neutrino and antineutrino collisions with
protons, deuterons, and Neon nuclei. 

     We analyze the data collected by
several neutrino experiments with big bubble chambers---BEBC at CERN
and the 15-foot chamber at Fermilab. These two bubble chambers were
close to each other in geometry, fiducial volume, and operating
conditions, and their data were collected and processed using very
similar techniques and algorithms. In the past, this already allowed to 
combine the neutrino data collected with BEBC and the 15-foot bubble 
chamber for a number of physics analyses \cite{combine}. A database
compiled by one of us (A.A.) comprises some 120 000 $\nu_\mu$- and
$\bar{\nu}_\mu$-induced charged-current events, and embraces the bulk
of neutrino data collected with BEBC (expts. WA21, WA25, and WA59) and
a significant fraction of data collected with the 15-foot bubble 
chamber\footnote{Unfortunately, our 
database does not include the biggest 
neutrino sample from the 15-foot bubble 
chamber---that collected by the $\nu$Ne 
experiment E53 \cite{baltay}.}
(expts. E180 and E632). Though obtained several decades ago, 
the neutrino data from big bubble chambers are still unrivaled in 
quality and completeness of physics information.

     In the BEBC experiments WA21 (hydrogen fill), WA25 (deuterium 
fill), and WA59 (neon-hydrogen mix), the data were collected using
essentially the same wide-band horn-focused beam, with mean energies
of \numucc\ and \anumucc\ events near 50 and 40 GeV, respectively.
The experiment E180 used the 15-foot bubble chamber filled with a
Ne--H$_2$ mix and exposed to a wide-band antineutrino beam under 
conditions very similar to WA59. In the last bubble-chamber experiment,
E632 at Fermilab, the 15-foot chamber was filled with a (lighter)
Ne--H$_2$ mix and exposed to a neutrino beam with quadrupole-triplet 
focusing from the Tevatron. In E632, mean energies of neutrino and 
antineutrino events reached some 140 and 110 GeV, respectively. 
Neutral-current interactions are not systematically included in the
database\footnote {In WA59, the bulk of 
NC events were rejected at scanning stage.},
and therefore our analysis is restricted to charged-current events
with $p_\mu > 4$ GeV. Total numbers and mean energies of \numucc\ and 
\anumucc\ events collected by the aforementioned experiments are 
summarized in Table \ref{statistics}. Further details on these neutrino 
experiments can be found in \cite{experiments}.

\begin{figure}[h]
\vspace{18 cm}
\includegraphics{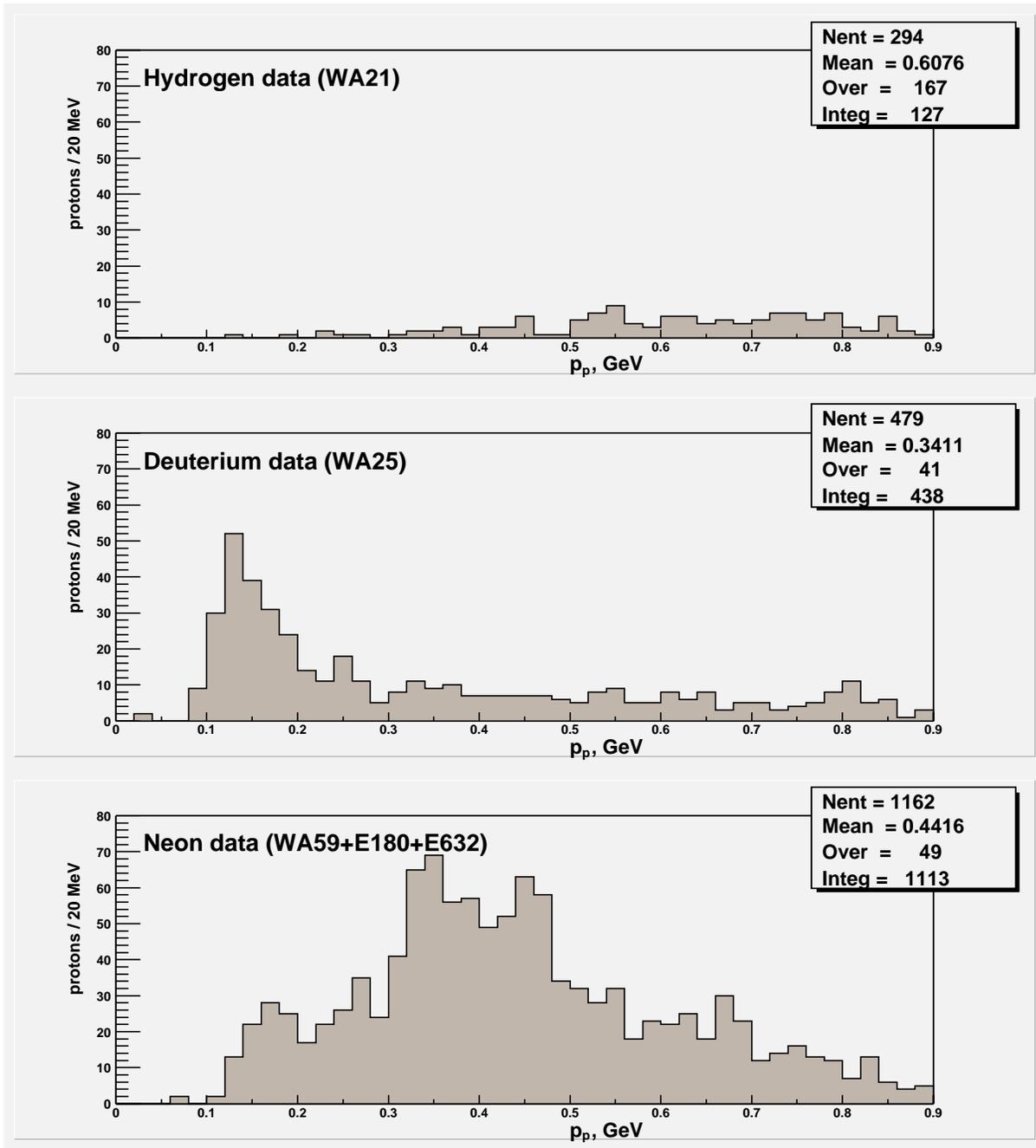}
\caption
{Momenta of identified protons emitted in association with $K^0_S$
mesons in \numucc\ and \anumucc\ collisions with Hydrogen, 
Deuterium, and Neon (top, middle, and bottom panels, respectively).}
\label{proton-momentum}
\end{figure}

     Unlike charged kaons that are virtually indistinguishable from
pions, neutral kaons are identified in a bubble chamber by 
reconstructing the decays \kpipi. On average, the $K^0(\bar{K^0})$
detection efficiency is near 25\%. At the same time, protons with
momenta below $\sim$ 1 GeV can
be identified by the stopping signature, bubble density, and variation 
of track curvature in magnetic field. Therefore, the $K^0 p$ channel 
seems mandatory when searching for formation of the $\Theta^+$ in a 
bubble chamber. The numbers of events featuring either reconstructed 
\kpipi\ decays and identified protons with $p_p < 900$ MeV, that are
used in this analysis, are quoted in Table \ref{statistics} for each 
(anti)neutrino sample considered. The momenta of protons identified in 
hydrogen, deuterium, and neon are plotted in Fig.~\ref{proton-momentum}. 
In deuterium, the enhancement at proton momenta below some 200 MeV is 
due to spectator protons. For neon events with reconstructed $K^0_S$ 
mesons and identified protons, mean proton multiplicity is $\simeq 1.4$.

\begin{figure}[h]
\vspace{18 cm}
\includegraphics{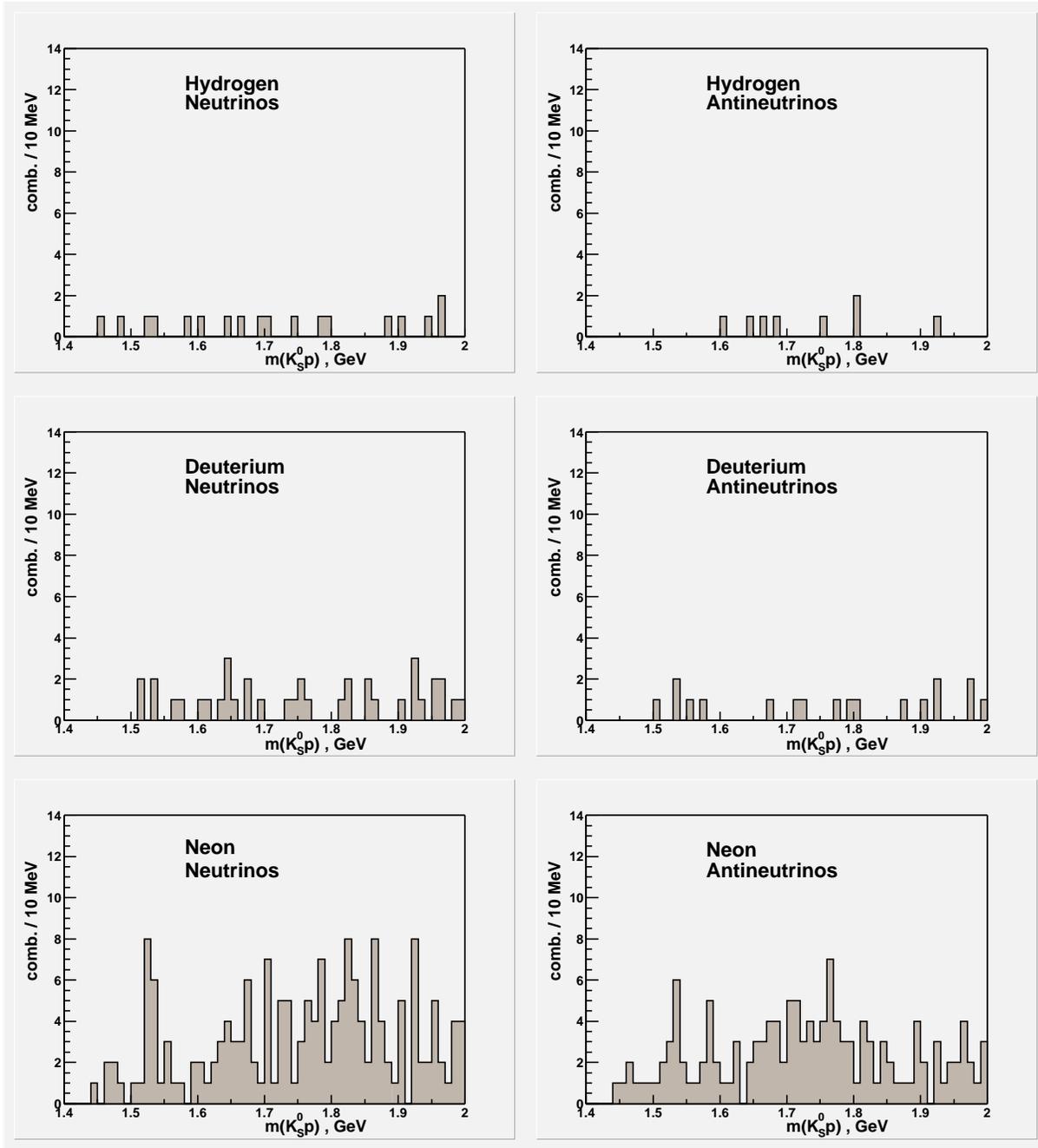}
\caption
{Invariant mass of the \kzepro\ system formed in \numucc\ (on the left) 
and \anumucc\ (on the right) collisions with Hydrogen, Deuterium, and 
Neon (top, middle, and bottom panels, respectively).}
\label{nu-anu}
\end{figure}

\begin{figure}[h]
\vspace{18 cm}
\includegraphics{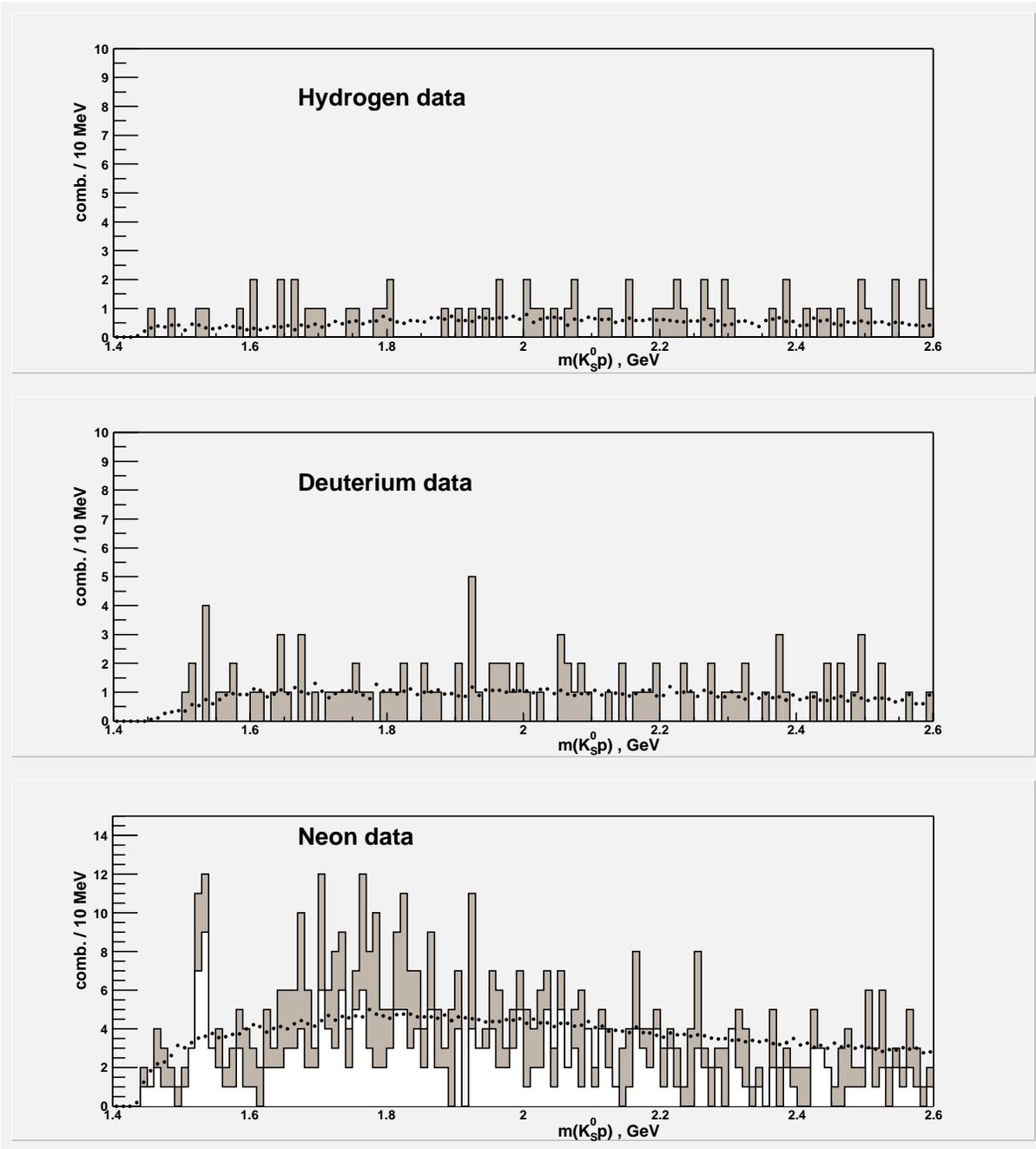}
\caption
{Invariant mass of the \kzepro\ system for the \numucc\ and \anumucc\ 
events combined. The top, middle, and bottom panels are for the
Hydrogen, Deuterium, and Neon data, respectively. The "random-star"
background obtained by pairing a $K^0_S$ from one event and a proton
from another event in depicted by dots. Dropping the events in Neon that 
feature two or more identified protons with $300 < p_p < 900$ MeV
results in the lower (open) histogram in the bottom panel.}
\label{combine-plus-rando}
\end{figure}

     For either fill, the \mkzepro\ distributions of \numucc\ and
\anumucc\ events are separately plotted in Fig. \ref{nu-anu} and 
combined---in Fig. \ref{combine-plus-rando}. Protons are selected
in the momentum interval of $300 < p_p < 900$ MeV. The combined 
$\nu + \bar{\nu}$ distribution for neon shows a distinct enhancement
at \mkzepro\ $\simeq 1530$ MeV. No neutrino events contribute twice or 
more to the peak region. The peak survives dropping the events that 
feature two or more identified protons with $300 < p_p < 900$ MeV, see 
the lower (open) histogram in the bottom panel of 
Fig. \ref{combine-plus-rando}. The combined $\nu + \bar{\nu}$ distribution 
for deuterium is also slightly enhanced in the same mass region.
The background in the peak region is estimated by pairing a $K^0_S$ from 
one event and a proton---from another event randomly selected in the same 
$\nu / \bar{\nu}$ subsample. Thus obtained "random-star" distribution
is then normalized to the \kzepro\ mass spectrum by the number of 
entries in the nonresonant region of \mkzepro~$> 2$ GeV (see the dotted 
histograms in Fig. \ref{combine-plus-rando}). It is noteworthy that, 
apart from the peak near 1530 MeV in the \mkzepro\ distribution for neon, 
the random-star background fails to reproduce a broad enhancement in the 
mass region 1650 $<$ \mkzepro\ $<$ 1850 MeV of the same spectrum. The
latter enhancement may be due to $\bar{K^0}p$ decays of a number of 
excited $\Sigma^{*+}$ states that populate this mass region \cite{pdg}.

\begin{figure}[h]
\vspace{18 cm}
\includegraphics{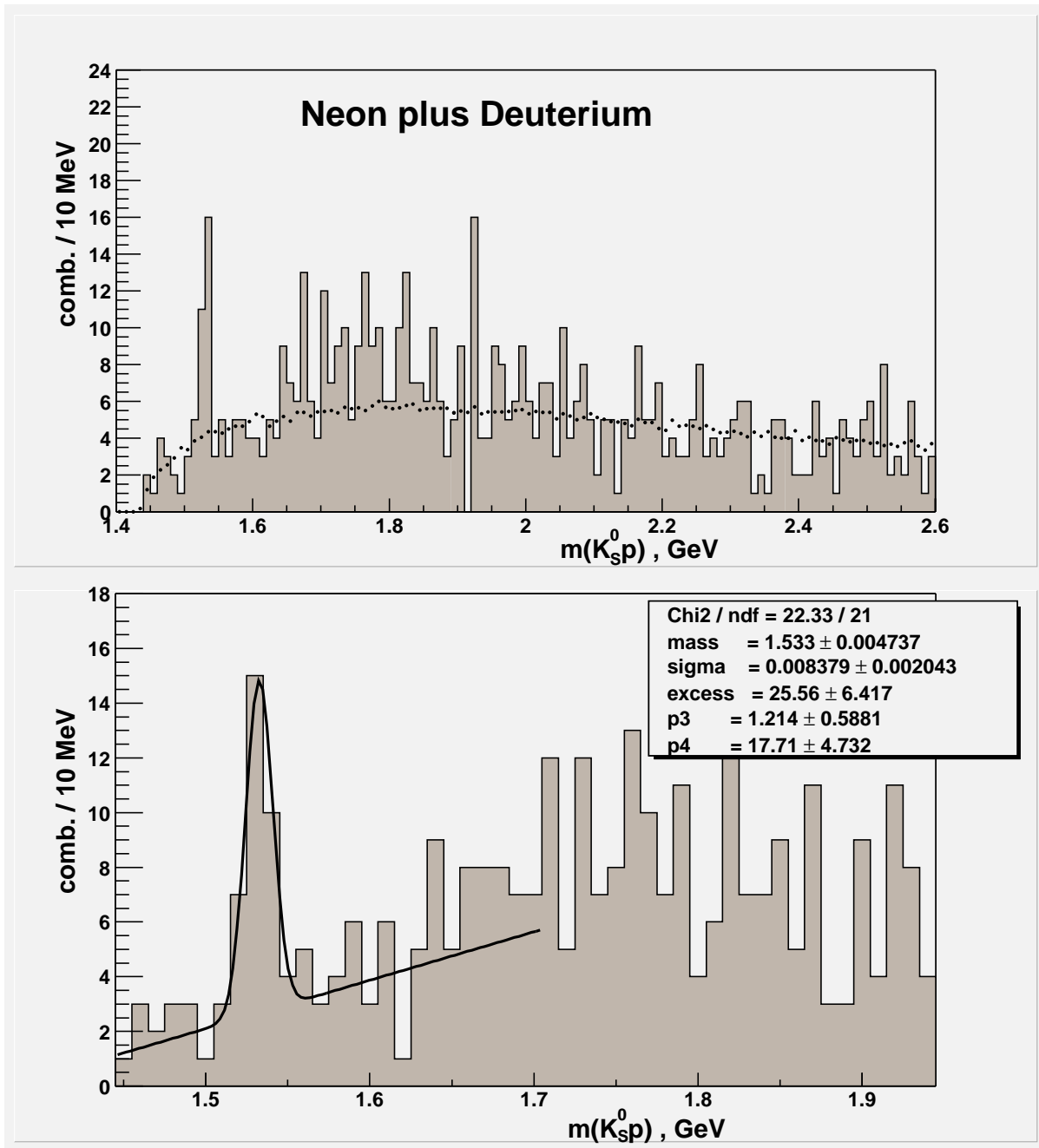}
\caption
{Invariant mass of the \kzepro\ system for the Neon and Deuterium data 
combined (top panel). The dots depict the random-star background.
A fit of the same \mkzepro\ distribution but plotted with shifted bins 
is shown in the bottom panel.}
\label{signal-two}
\end{figure}

     Figure \ref{signal-two}  shows the \mkzepro\ distribution
for the Neon and Deuterium data combined. In two 10-MeV bins between
1520 and 1540 MeV, we have 27 events with a background of $\sim 8$ events
as estimated from random \kzepro\ pairs (see the dotted histogram). The 
statistical significance of the peak is thus near 6.7 standard 
deviations. It makes no sense to fit a signal restricted to just two 
bins as in the top panel of Fig. \ref{signal-two}, so in the bottom
panel we plot the same \mkzepro\ distribution with bins shifted by 5 MeV.
A fit of the latter histogram to a Gaussian on top of linear background
returns $M = 1533 \pm 5$ MeV and $\sigma = 8.4 \pm 2.0$ MeV for the
position and r.m.s. width of the resonance, respectively. The r.m.s.
width is found to be consistent with experimental resolution on \mkzepro\
estimated from live events in the peak ($\simeq 8.5$ MeV). 

     For neutrino and antineutrino events that contribute to the peak
region of 1510 $<$ \mkzepro\ $<$ 1550 MeV, mean values of $E_\nu$
($57 \pm 10$ GeV) and $Q^2$ ($12.5 \pm 3.3$ GeV$^2$) are consistent
with those for all CC events with detected $K^0_S$ mesons. Mean
momentum of the \kzepro\ system for peak events, 
$\langle p(K^0_S p) \rangle = 1.08 \pm 0.06$ GeV, is much less than
that of all detected $K^0_S$ mesons (see Table \ref{statistics}).
%  The smallness of \kzepro\ momentum does not necessarily reflect the
%  true dynamics of the process: a soft identified proton has to couple
%  with an equally soft $K^0_S$ to form a \kzepro\ system with mass near
%  threshold.

     Unfortunately, neutrino data do not allow to determine the 
strangeness of the observed resonant state with mass near 1533 MeV, as 
was done in \cite{nakano, xenon, clas, saphir}. On the other hand, 
there are no known $\Sigma^{*+}$ states in this mass region. Therefore,
we interpret the enhancement near 1533 MeV observed in the \mkzepro\ 
distribution as a signal from formation of the $\Theta^+$ baryon in 
neutrino and antineutrino collisions with nuclei. The mass and width of 
this state are estimated as $M = 1533 \pm 5$ MeV and $\Gamma < 20$ MeV,
respectively. The cross section of $\Theta^+$ production by neutrinos
appears to increase with atomic number of the target nucleus.

     The excellent neutrino data analyzed in this paper are a result
of painstaking and ingenious work of the WA21, WA25, WA59, E180, and E638
experimental teams over many years, and their efforts are gratefully 
acknowledged. We wish to thank Prof. M.V. Danilov, Prof. L.B. Okun, 
Dr. V.S. Borisov, and Dr. V.S. Verebryusov for useful comments and 
suggestions.

\clearpage

\begin{table}[h]
\begin{tabular}{|l|c|c|c|c|c|}
\hline
Experiment                  &  WA21 &  WA25 &  WA59 &  E180 & E632 \\
Chamber                     &BEBC   &BEBC   &BEBC &15' B.C.&15' B.C. \\
Fill      &Hydrogen &Deuterium &Neon--H$_2$ &Neon--H$_2$ &Neon--H$_2$\\
\hline
\hline
{\bf Neutrinos:}             &       &       &       &       &     \\
Mean $E_\nu$, GeV            &48.8   &51.8   &56.8   &52.2   &136.8 \\
Mean momentum of             &5.7    &5.7    &4.5    &3.4    &7.7   \\
\ \ \ \ \ detected $K^0_S$, GeV&       &       &       &       &       \\
All measured CC events       &18746  &26323  & 9753  & 882   & 5621  \\
CC events with $K^0_S$       & 1050  & 1279  &  561  &  21   &  587 \\
CC events with $K^0_S$ and &82 (78)&307 (128)&193 (193)& 9 (8)&229 (157) \\
\ \ \ \ \ identified protons     &       &       &       &       &      \\
\hline
{\bf Antineutrinos:}         &       &       &       &       &       \\
Mean $E_\nu$, GeV            &37.5  &37.9    &39.5   &33.8 &110.0 \\
Mean momentum of             &4.2    &4.2    &3.5     &3.4    &7.6    \\
\ \ \ \ \ detected $K^0_S$, GeV &       &       &       &       &       \\
All measured CC events      &13155  &16314  &15693  & 5927  &1190   \\ 
CC events with $K^0_S$       &  702  &  761  &  631  &  231  & 123   \\
CC events with $K^0_S$ and &45 (43)&116 (57)&185 (185)&56 (54)&49 (28)   \\
\ \ \ \ \ identified protons &       &       &       &       &      \\
\hline
\end{tabular}

\caption
{For the five neutrino experiments considered, mean energies of
$\nu_\mu$- and $\bar{\nu}_\mu$-induced CC events, mean momenta of 
$K^0_S$ mesons reconstructed by \kpipi\ decays,  and the numbers of all 
measured CC events, of those with detected $K^0_S$ mesons, and of those 
featuring either $K^0_S$ mesons and identified protons with momentum 
$p_p < 900$ MeV. The numbers in parentheses are for the additional
selection of $p_p > 300$ MeV. Note that in the experiment E632, all
neutrino events were measured on part of exposed film, and only those
that showed \kpipi\ and $\Lambda^0 \rightarrow p \pi^+$ candidates
were measured on another part of the film.}
\label{statistics}
\end{table}

\end{document}